\documentclass[a4paper,12pt]{revtex4}
\usepackage{amssymb}
\usepackage{amsmath}
\usepackage{graphics}
\usepackage{color}
\usepackage[T1]{fontenc}
\usepackage{times,mathptm}
\usepackage[cp1250]{inputenc}
\usepackage{epsfig,epsf}
\newcommand{\be}{\begin{equation}}
\newcommand{\ee}{\end{equation}}

\newcommand\beq{\begin{eqnarray}}
\newcommand\eeq{\end{eqnarray}}
\begin{document}

\title{Chiral density waves in quarkyonic matter}

\author{Tomasz L. Partyka and Mariusz Sadzikowski}

\affiliation{Smoluchowski Institute of Physics, Jagellonian
University, Reymonta 4, 30-059 Krak\'ow, Poland}

\begin{abstract}
We study the phase diagram of strongly interacting matter including
the inhomogeneous phase of chiral density waves (CDW) within the
Polyakov loop extended Nambu - Jona-Lasinio (PNJL) model. We discuss
the phase structure taking into account density and flavour
dependence of the Polyakov loop potential parameter and temperature
dependence of the four-point coupling constant of the NJL model.
It is shown that the CDW phase exists and that can be interpreted as
a special realisation of quarkyonic matter. This fact is of
particular interest because the existence of homogeneous quarkyonic
matter is strongly constrained. This also indicates that the study
of inhomogeneous phases at finite temperatures and baryon densities
are of special importance.
\end{abstract}


\maketitle

\section{Introduction}


An understanding of structure of the phase diagram of strongly interacting matter at finite baryon density is still
an open problem despite of the many efforts that have been devoted to its study since the very
beginning of the QCD era. The lack of experimental data and reliable tools of the
direct QCD based calculations are the main culprits of this situation.
Nevertheless, the progress is still possible and our experience must rely on the interplay
between general arguments, models calculations and QCD lattice results accessible
at the low chemical potential. A very good example
of such approach is the PNJL model \cite{fukushima} which combines the
chiral and the deconfinement order parameters linked to QCD at finite temperature
through the parameter fits to the lattice data \cite{ratti,rosner,ratti2}.

The new insight into the high baryon density domain based on large $N_c$ expansion
has been proposed by McLerran and Pisarski \cite{mclerran}. In the limit of a large
number of colours $N_c\rightarrow\infty$ the QCD phase diagram simplifies substantially. The low baryon density and temperature
region is occupied by the confined, chirally broken phase where the matter pressure scales with $N_c$ as $P\sim O(1)$.
The high temperature phase is deconfined and dominated by gluons with the $P\sim N_c^2$ scaling. Finally,
at the low temperature and high baryon density there is a confined phase which scales as $P\sim N_c$. The
latter phase has been named quarkyonic.

It is very natural to interpret the quarkyonic phase within the
PNJL model as deconfined and chirally restored phase. This is based
on the PNJL picture of the dense matter which consists of a degenerate Fermi sea of quarks with the
colorless particle-hole excitations at the Fermi surface. Then the $P\sim N_c$  scaling at high density
naturally arises. It was shown in
paper \cite{mclerran_redlich} that at  $N_c=3$ the quarkyonic phase slightly precedes
the chiral transition at finite density and low temperature. The existence of the quarkyonic phase within the PNJL model was also studied in Ref. \cite{Abuki}.

Discussions of the QCD phase diagram frequently consider only homogeneous phases.
However, at the high baryon density one also has to take into account the possibility
of the crystal structures. The very well known examples are large $N_c$ QCD \cite{deryagin},
skyrmion crystals \cite{skyrmion}, LOFF phases \cite{LOFF}, Overhauser effect \cite{overhauser} and
chiral density waves \cite{CDW}. The competition between chiral
density waves and uniform colour superconductors were also discussed \cite{MS}.
The chiral spirals has been already considered in the quarkyonic phase \cite{kojo}. In this letter we
would like to consider the chiral density waves in quarkyonic matter from the
point of view of the PNJL model. It is an interesting problem to check how
the chiral density waves are influenced by the presence of the Polyakov loop field
and what kind of the feedback it generates. We also pay a particular attention to
the relation between the inhomogeneous chiral phase and the quarkyonic matter.

\section{Chiral density waves in the PNJL model}

We consider the PNJL model with two light quarks and three colours with
the parametrisation of the Polyakov loop based on paper \cite{rosner} and the NJL part
based on paper \cite{klevansky}
\beq
\mathcal{L}=\bar{\psi}i\gamma^\mu D_\mu\psi+G[(\bar{\psi}\psi)^2+(\bar{\psi}i\gamma_5\vec{\tau}\psi)^2]-U(\Phi,T)
\eeq
where $\psi$ is a massless quark field and $D_\mu = \partial_\mu-iA_\mu$ is covariant derivative.
The $SU(3)$ gauge field $A_\mu = (A_0,\vec{0})$, $A_0=gA^a_0\lambda_a/2$ and $\lambda_a$ are the
Gell-Mann matrices. The effective potential describing the traced Polyakov loop
$$
\Phi = \frac{1}{N_c}\mbox{Tr}\left[\mathcal{P}\exp\left(i\int_0^\beta d\tau A_4\right)\right],
$$
in the fundamental representation takes the form \cite{rosner}
\beq
\label{poly_loop_pot}
\frac{U(\Phi,T)}{T^4}&=&-\frac{1}{2}a(T)\Phi^2+b(T)\ln [1-6\Phi^2+8\Phi^3-3\Phi^4], \nonumber\\
a(T) &=& 3.51 - 2.47 \frac{T_0}{T} + 15.2 \left(\frac{T_0}{T}\right)^2,\;\;\; b(T) = -1.75 \left(\frac{T_0}{T}\right)^3,
\eeq
where $A_4=iA_0$ is a colour gauge field and the temperature $T_0=270$ MeV
describes the deconfinement transition in a pure gauge sector.
In equation (\ref{poly_loop_pot}) we already used the constraint that the expectation value of the Polyakov loops
$\langle\Phi\rangle$, $\langle\Phi^\ast\rangle$ are real \cite{dumitru} which gives $\Phi=\Phi^\ast$
at the mean field level \cite{rosner,ratti2}. Above treatment may be improved with fluctuations that lead to a difference between the expectation values of the traced Polyakov loops at non-zero baryon densities \cite{Hell}. However, in the present work, as a practical procedure, we consistently keep $\Phi=\Phi^\ast$.
Among the literature, one encounter also another approach. Fields $\Phi$ and $\Phi^{*}$ are treated as independent in the minimalization of the grand thermodynamic potential \cite{sasaki}.

The coupling constant $G=5.024$ GeV$^{-2}$ and the three dimensional momentum cut-off $\Lambda = 0.653$ GeV was used for regularization of the divergent vacuum contribution which reproduce the correct values of the pion decay constant and chiral condensate.

We are working at the mean field level with the chiral density wave ansatz \cite{CDW}
\be
\label{ansatz}
\langle\bar{\psi}\psi\rangle = M\cos \vec{q}\cdot \vec{x},\;\;\;
\langle\bar{\psi}i\gamma_5\tau_3\psi\rangle = M\sin \vec{q}\cdot \vec{x},
\ee
where the third direction in isospin space was chosen arbitrarily. The colour field
in a Polyakov gauge creates a constant background field $A_4=\lambda_3\phi$ which is related
to the Polyakov loop through the equation $\Phi = (1+2\cos\phi)/3$.

The standard procedure of the mean-field calculation
in the Matsubara formalism leads to the thermodynamic potential of the form
\beq
\frac{\Omega}{V} = \frac{M^2}{4G}+U(\Phi,T)-\frac{T}{2}\sum_n\int\frac{d^3p}{(2\pi)^3}
\mbox{Tr}\ln[S^{-1}(i\omega_n,\vec{p}]
\eeq
where $\omega_n=(2n+1)\pi T$. The inverse propagator in Nambu-Gorkov space reads
\beq
S^{-1}(i\omega_n,\vec{p}) =
\left[
\begin{array}{cc}
i\gamma_0(\omega_n+A_4+i\mu)+\vec{\gamma} \left(\vec{p}-\frac{1}{2}\gamma_5\tau_3\vec{q}\right)+M & 0 \\
0 & i\gamma_0(\omega_n-A_4-i\mu)+\vec{\gamma} \left(\vec{p}-\frac{1}{2}\gamma_5\tau_3\vec{q}\right)+M
\end{array}
\right]
\eeq
where $\mu$ is a quark chemical potential.
After diagonalization of the propagator one finally arrive at the formulae
\beq
\label{potential}
\frac{\Omega}{V} &=& U(\Phi,T)+\frac{M^2}{4G}+\frac{M^2F_\pi^2\vec{q}^2}{2M_0^2}-12\int^\Lambda\frac{d^3p}{(2\pi)^3}E_0 - \nonumber\\
&-& 2T\sum_{i=\pm}\int^\Lambda\frac{d^3p}{(2\pi)^3}\left\{\ln\left[1+3\Phi e^{-(E_i-\mu)/T}\left(1+e^{-(E_i-\mu)/T}\right)
+e^{-3(E_i-\mu)/T}\right]\right.\nonumber\\
&+&\left.\ln\left[1+3\Phi e^{-(E_i+\mu)/T}\left(1+e^{-(E_i+\mu)/T}\right)
+e^{-3(E_i+\mu)/T}\right]\right\}
\eeq
where $M_0=0.301$ GeV is a constituent quark mass at zero temperature and density. The regularization through the
3-dim momentum cut-off was introduced after paper \cite{MS}. The energy eigenvalues are given
by the expressions
$$
E_\pm = \sqrt{\vec{p}^2+M^2+\frac{\vec{q}^{\,2}}{4}\pm \sqrt{(\vec{q}\cdot\vec{p})^2+M^2\vec{q}^{\,2}}},\;\;\;
E_0 = \sqrt{\vec{p}^2+M^2}.
$$
Let us notice that potential (\ref{potential}) reduces to NJL model prediction in the deconfined limit $\Phi=1$ .

\section{The phase diagram}

The global minima of the potential (\ref{potential}) as a function of temperature and baryon density
describe the phase diagram of the strongly interacting matter. This prescription
leads to the self-consistent equations
$$
\frac{\partial\Omega}{\partial M} = \frac{\partial\Omega}{\partial \Phi} = \frac{\partial\Omega}{\partial |\vec{q}|} = 0.
$$

\begin{figure}[h]
\centerline{\epsfxsize=8.5 cm \epsfbox{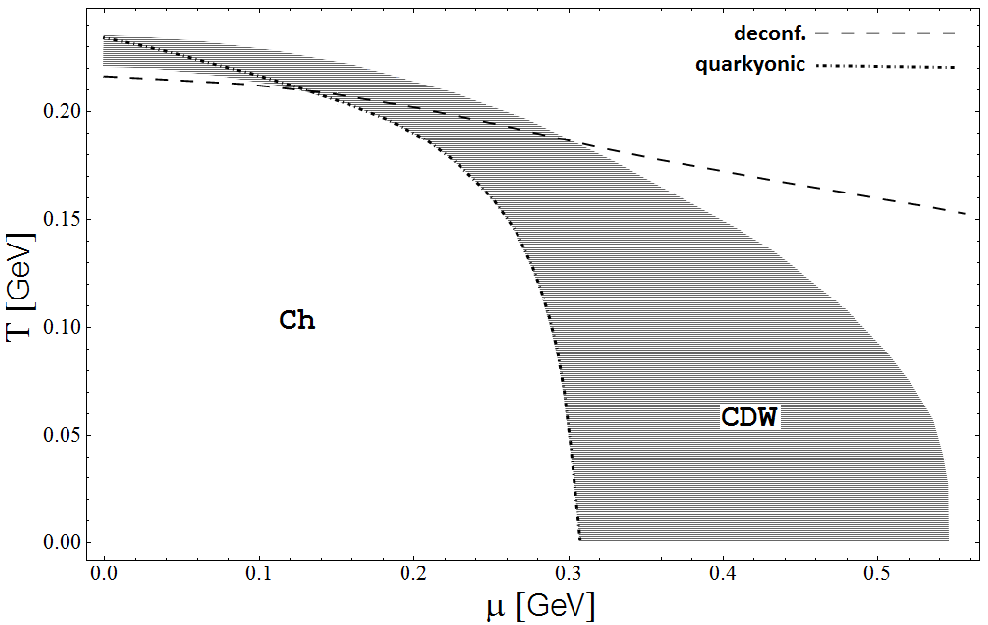} \epsfxsize=8.5 cm \epsfbox{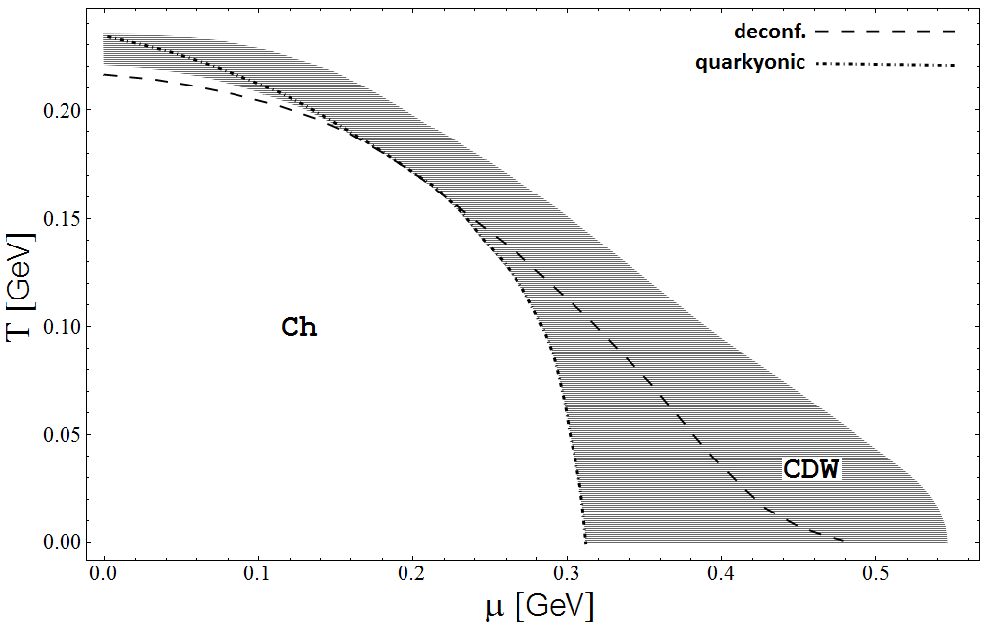}}
\caption{The phase diagram of the PNJL model with inhomogeneous chiral wave. The dashed curves
are lines of the deconfinement phase transition. The left panel shows the diagram
for the constant Polyakov loop potential parameter $T_0=270$ MeV, whereas the right panel
for the density dependent parameter $T_0(\mu)$ given by equation (\ref{T0(mu)}).
The gray colour describes the inhomogeneous CDW phase. The dotted curves are lines of
the transition to quarkyonic phase. These lines essentially coincide
with the lines of the first order phase transition to CDW phase at higher density.}
\end{figure}

Fig. 1 shows the the PNJL model phase diagram including the inhomogeneous chiral
density wave phase (CDW) marked with the gray colour. This phase is surrounded by the line of the first order
phase transition where there are jumps in the values of the all order parameters $M, \Phi$ and $|\vec{q}|$.
The jump in $\Phi$ is induced by the first order phase transition in $M$ and it is not a mark
of the transition to a deconfinement phase. One expects rather a crossover here and the exact
place of this transition is not fixed unambiguously. We define the transition line as
a place where the derivative $d\Phi/dT$ reaches its maximum unless stated otherwise
(see subsection IIIB). Using our definition the lines of deconfinement phase transition are shown in Fig. 1
as dashed curves. For the Polyakov loop potential parameter $T_0=270$ MeV
the chiral and deconfinement transitions almost perfectly coincide at zero density,
however, they split around the point $(T,\mu)=(0.21,0.12)$ GeV
(left panel of Fig. 1). It is interesting to note that this is
also the place where the quarkyonic phase appears. The boarder line for the quarkyonic matter is defined as a place where the value of the quark chemical potential exceeds the value of the constituent quark mass \cite{mclerran_redlich}. 
This line perfectly coincides with the first order phase
transition line of the CDW phase for $\mu>0.12$ GeV. It happens because at the first order
phase transition there is a large jump
in the value of the constituent quark mass $M$ which drops below the value of $\mu$.
One can conclude that within the PNJL model the quarkyonic matter can be treated
as confined, however, spatially inhomogeneous phase.

\subsection{Deconfinement and the baryon density}

It is important to remember that the critical temperature of the deconfinement phase transition
decreases with increasing baryon number density. This results in $T_0$ parameter dependence on the
quark chemical potential \cite{schaefer}. It can be intuitively understand in the picture
of overlaping hadrons. At zero density the finite temperature causes fluctuations of mesons
and baryon-antibaryon pairs up to some critical value of $T_c$. Above this temperature hadrons
overlap with each other and there is a possibility of a colour flow
in space which one can interpret as the process of deconfinement.
At higher baryon chemical potential there are already some number of hadrons present in a medium. Then
a new critical temperature, lower then $T_c$, is sufficient to create appropriate number of
hadrons which start to overlap. In paper \cite{fukushima2} K. Fukushima tried to estimate
such dependence from the statistical model data which we also adopted here
\be
\label{T0(mu)}
T_0(\mu)=T_0-9 b \mu^2,
\ee
where $b=1.39\cdot 10^{-4}$ MeV$^{-1}$ and for $T_0(\mu)<1$ MeV the phase is
defined as deconfined. On the right panel of Fig. 1 the dashed curve describes
the deconfinement phase transition for the density dependent parameter (\ref{T0(mu)}). As expected
the line bends toward the density axis, but still, there is a place for the inhomogeneous phase in the
quarkyonic region for temperatures below 0.16 GeV and chemical potential above 0.22 GeV.
This needs to be confronted with the statement that the existence of homogeneous quarkyonic
phase is inconsistent with the prediction of statistical model \cite{fukushima2}.
The introduction of non-uniform phases open up a window for realisation
of the quarkyonic matter (see also Fig. 4).

\subsection{Deconfinement and the number of flavours}

A value of the temperature parameter $T_0$ depends also on the number of
active flavours. For the model which contains two degenerate flavours
one should reduce the temperature parameter to  $T_0=208$ MeV
\cite{ratti,schaefer}. The phase diagram for this parameter is shown on the left panel of Fig. 2.
First of all one looses a good coincidence between the chiral and
deconfinement phase transitions at zero density, nevertheless, the CDW phase remains
intact. The confined (quarkyonic), inhomogeneous phase appears
at chemical potential larger then $\mu=0.21$ GeV and temperatures below $T=0.16$ GeV.
For lower baryon densities and higher temperatures the CDW phase still exists but it is deconfined.

\begin{figure}[h]
\centerline{\epsfxsize=8.5 cm \epsfbox{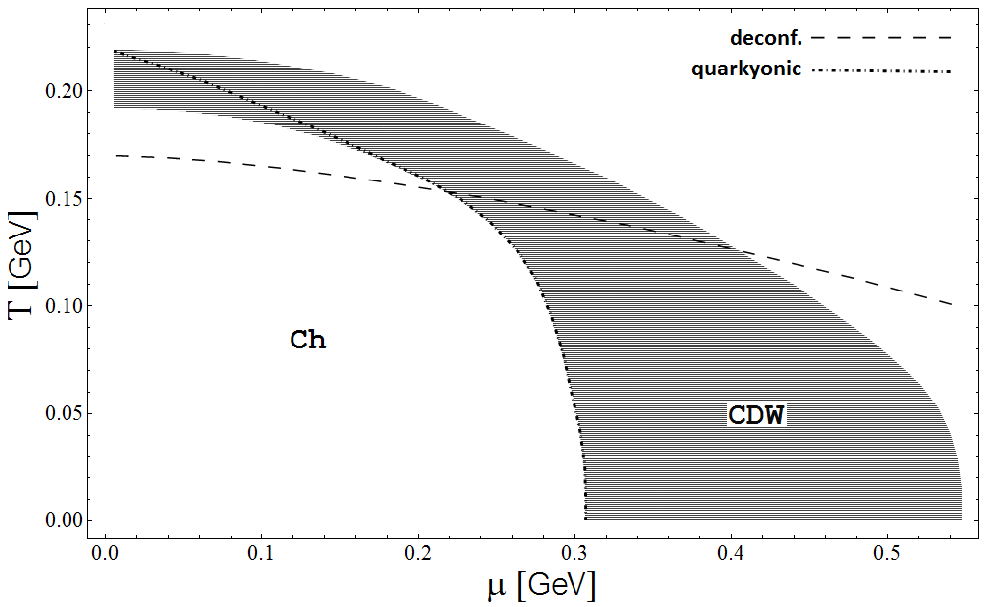} \epsfxsize=8.5 cm \epsfbox{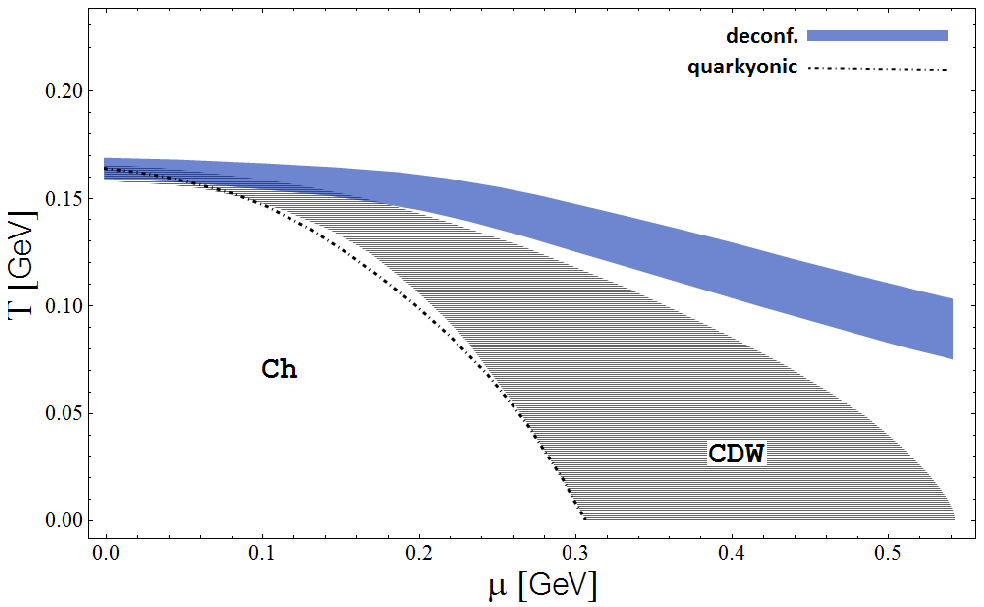}}
\caption{The phase diagram of the PNJL
model with inhomogeneous chiral wave for the Polyakov loop potential
parameter $T_0=208$ MeV which corresponds to $N_f=2$ active
degenerate flavours. On the left panel the NJL parameter is set to
$G=5.024$ GeV$^{-2}$. On the right panel the temperature dependence
of the NJL parameter $G(T)$ is taken into account according to
equation (\ref{GT}). Behavior of the Polyakov loop is disturbed by the first order transitions into the inhomogeneous phase and consequently deconfinement transition is determined with uncertainty within the blue band.}
\end{figure}

The phase diagram at $T_0=208$ MeV is not satisfactory in a sense
that at zero density there is a mismatch between the chiral and
deconfinement phase transitions which contradicts the lattice
results \cite{lattice}. One can consider a possibility that this
problem is a consequence of the wrong value of the coupling constant
$G$ at non-zero temperature. The effective coupling constant $G$ is
in principle a function of temperature and density. It contains
contributions which follow from the integration of the gluonic
degrees of freedom. At higher values of $T$ gluons interaction with
quarks weakens which in turn influences the effective four-quark
interaction in the same way. Thus the coupling $G$ decreases with
increasing temperature which lowers the critical temperature of the
chiral phase transition. If one attributes the whole mismatch
between the chiral and deconfinement transitions at zero density to
the wrong value of $G$ then one can try, at least at the
phenomenological level, to reestablish the agreement changing its
value into the new one $G(T_c)=G_c$ in such a way, that both
transitions has the same critical temperature $T_c$. We assume here
that the coupling $G$ is a linear function of temperature
\beq
\label{GT} &&G(T)=G(1-(T/T_c))+G_c(T/T_c),\\\nonumber
&&G=5.024\,\mbox{GeV}^{-2},\;\; G_c=4.221\,\mbox{GeV}^{-2},\;\;T_c=167\,\mbox{MeV}.
\eeq
According
to our fit the change in the constant $G$ between zero temperature
and $T_c$ is of the order of 15 per cent which is not much. Then one
can treat equation (\ref{GT}) as a series expansion in the
temperature around $T=0$. The density dependence of $G$ is neglected
since it is less important at large $N_c$ limit where quarks
decoupled from the gluonic degrees of freedom.

The right panel of Fig.2 describes the phase diagram where the NJL coupling constant $G$
is given by the linear function from eq. (\ref{GT}). It is clearly seen that the chiral and
deconfinement phase transitions coincides at zero density to a good approximation.
Let us mention that the usual definition of the deconfinement transition line as
a place where the derivative $d\Phi/dT$ reaches its maximum is not quite useful
in a situation where $\Phi$ is a discontinues function of temperature and
the points of discontinuity are close to the expected maximum value of $d\Phi/dT$.
In such situations we define the deconfinement
transition as a place where the Polyakov loop $\Phi$ takes the value $0.31\pm 0.07$.
This is rather a modest value, however, such range is suggested when
one considers the deconfinement phase transition for the homogeneous phases
where the transition line is defined in a standard way.
Using our definition the line of deconfinement phase transition is shown in Fig. 2
(later in Fig. 4) as a dark (blue) band.
At low temperature and high baryon density one recovers the results of the NJL model
with a constant $G$ coupling.

\begin{figure}[h]
\centerline{\epsfxsize=8.5 cm \epsfbox{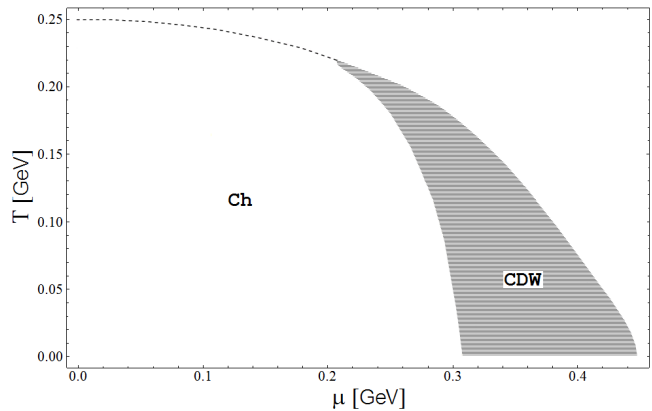}}
\caption{The chiral phases of the PNJL model with inhomogeneous chiral wave. The
NJL model parameters are $G=5.024$ GeV$^{-2}$ and $\Lambda = 0.653$ GeV. The temperature dependent
part of the potential (\ref{potential}) is also regularized by the finite cut-off. In such a regularization
the CDW phase is limited to the high density region of the diagram.}
\end{figure}

\subsection{Triple point}

The concept of a triple point in the QCD phase diagram refers to various situations.
For instance, the existence of a triple point between hadronic, color-superconducting and quark-gluon sectors was discussed in Ref. \cite{Blaschke}.
In the context of the present work, we refer to a very recent idea, namely, to the triple point in which hadronic, quarkyonic and quark-gluon phases meets together. Such a possibility was emphasized in paper \cite{Andronic}.
Although in Figs. 1,2 the CDW phase exists even at zero density region (there is no space for triple point), but the location of the inhomogeneous phase depends on the regularization
parameter $\Lambda$. In Figs. 1,2 the cut-off parameter in the temperature dependent
part of the potential (\ref{potential}) was sent to infinity. If one keeps
this parameter at the constant value $\Lambda = 0.653$ GeV then the phase diagram
changes and it is shown in Fig. 3. In this case the triple point actually appears at the point where the inhomogeneous phase ends $(T_3,\mu_3)=(0.22,0.2)$ GeV, what is in agreement with the results of Ref. \cite{Nickel}. However,
one has to remember that the location of this point is strongly dependent on the model
parameters and the regularization method.

For the quark chemical potential below
$\mu_3$ the phase line describes the continuous phase transition whereas above $\mu_3$ the line is of the first order.
When a non-zero current quark mass is turned on then the continuous phase
transition changes into a smooth crossover and only the island of CDW phase
remains on the diagram.

\section{Summary and conclusions}

We have discussed the phase diagram of the strongly
interacting matter including the spatially dependent chiral density waves in the PNJL model.
It was shown that the inhomogeneous CDW phase exists and dominates over the chirally restored phase
in a large domain of the phase diagram. We have also pointed out that the chiral density wave
can be interpreted as a special realisation of the quarkyonic matter.
This is an interesting possibility, particularly, when the homogeneous quarkyonic
phase was strongly constrained by the results of the statistical model \cite{fukushima2}.
Indeed in a case of two flavours the phase diagram with only homogeneous phases
is shown in Fig. 4 (left panel) where the effects of
flavour and density dependence of the Polyakov loop potential parameter $T_0(N_f,\mu)$
as well as the temperature dependence of the NJL coupling constant $G(T)$
were taken into account. This last dependence reflects the fact that at higher
temperature the four-point quark interaction should weaken which in turn let
the chiral and the deconfinement transitions stay at the same critical temperature
at zero baryon density. In the diagram the deconfinement phase transition precedes
the chiral phase transition almost in all the domain. Only at high density and
low temperature the transitions start to coincide. In such situation
there is no space left for quarkyonic matter in accordance with the conclusion
of paper \cite{fukushima2}.

\begin{figure}[h]
\centerline{\epsfxsize=8.5 cm \epsfbox{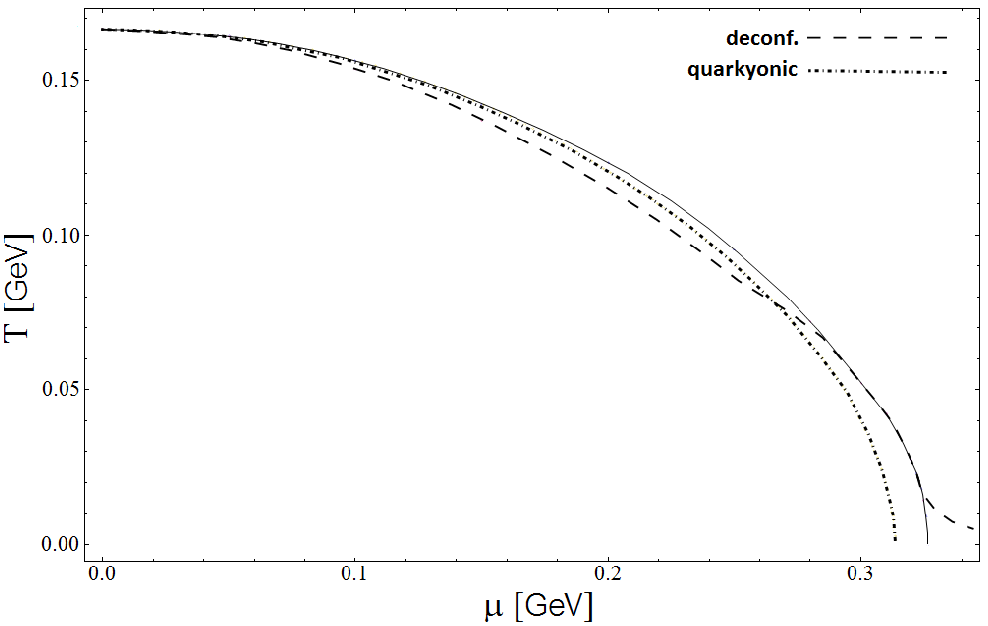} \epsfxsize=8.5 cm \epsfbox{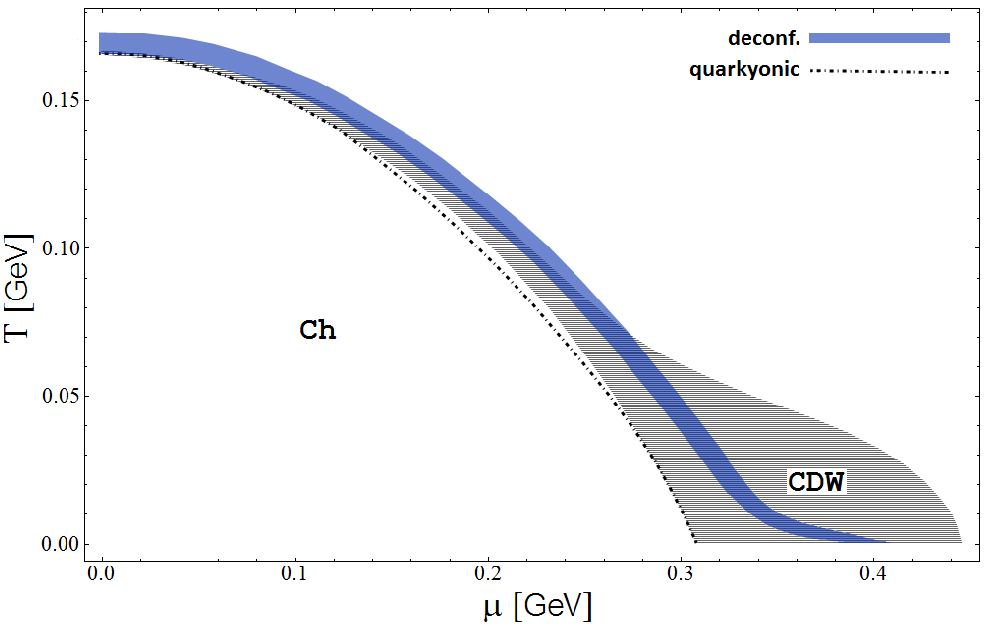}}
\caption{Left panel: the phase diagram of the PNJL model with homogeneous phases only. Above the point (T,\;$\mu$) = (0.085,\;0.263), the nature of the chiral transition changes from first to second oreder.
Right panel: the phase diagram of the PNJL model with inhomogeneous chiral density wave. The Polyakov
loop potential parameter $T_0(\mu)$ is given by (\ref{T0(mu)}) with $T_0=208$ MeV.
The finite 3-dim momentum cut-off $\Lambda=0.653$ GeV regularized the potential (\ref{potential})
and the NJL coupling constant $G(T)$ depends on temperature through the eq. (\ref{GT}).
The grey colour depicts the CDW phase. The dark (blue) band shows a location of the
deconfinement phase transition.}
\end{figure}

The phase diagram in the right panel of Fig. 4 sums up all physical effects that we discussed in the previous sections and is the best candidate for the QCD phase diagram. The left panel of Fig. 4 should be compared with the right panel of the same figure.
Let us notice a clear change in the
order of the phase transitions. The phase transition to the CDW phase precedes the deconfinement
phase transition. This fact also opens a window for the quarkyonic matter.
The transition into the quarkyonic phase is close to or coincides with the first order phase
transition into the CDW phase. Then the matter is spatially inhomogeneous but
still confined. Let us remind that the details of the phase diagram
depends on the temperature parametrisation of the NJL coupling constant $G(T)$. Nevertheless,
at low temperatures one should expect the same pattern of phase transitions as given in Fig. 4.

For the model parameters we choose the CDW phase
exists up to a zero density line. However, the low density and high temperature region
is strongly affected by the temperature fluctuations which are neglected
in the mean field approximation. These fluctuations probably melt
the "crystal" structure of the CDW phase and the triple point
is expected to appear on the phase diagram eventually.

It is an interesting task for the future work to compare the phases of the chiral density waves and the chiral spirals
\cite{kojo}. However, it requires the implementation of the chiral spirals within the NJL model in the
first place. This is a subject of certain
importance because the quarkyonic matter, if exists, would be most probably
of an inhomogeneous nature. One should also check the dependence of the results against the regularization scheme
and the influence of the finite current quark mass \cite{partyka}.
Finally, it is of great interest to consider more general ansatz then (\ref{ansatz}) to study
the possibility of the creation of different crystal structures.


\begin{thebibliography}{99}
\bibitem{fukushima} K. Fukushima, Phys. Lett. \textbf{B591} (2004) 277.
\bibitem{ratti} C. Ratti, M. A. Thaler and W. Weise, Phys. Rev. \textbf{D73} (2006) 014019.
\bibitem{rosner} S. R\"o{\ss}ner, C. Ratti and W. Weise, Phys. Rev. \textbf{D75} (2007) 034007.
\bibitem{ratti2} C. Ratti, S. R\"o{\ss}ner, M. A. Thaler and W. Weise, Eur. Phys. J. \textbf{C49} (2007) 213.
\bibitem{mclerran} L. McLerran and R. D. Pisarski, Nucl. Phys. \textbf{A796} (2007) 83.
\bibitem{mclerran_redlich} L. McLerran, K. Redlich and C. Sasaki, Nucl. Phys. \textbf{A824} (2009) 86.
\bibitem{Abuki} H. Abuki, R. Anglani, R. Gatto, G. Nardulli and M. Ruggieri, Phys. Rev. \textbf {D78} (2008) 034034.
\bibitem{deryagin} D. V. Deryagin, D. Y. Grigoriev and V. A. Rubakov, Intl. Jour. Mod. Phys. \textbf{A7} (1992) 659.
\bibitem{skyrmion} M. Kutschera, C. J. Pethick and D. G. Ravenhall, Phys. Rev. Lett \textbf{53} (1984) 1041;
I. R. Klebanov, Nucl. Phys. \textbf{B262} (1985) 133.
\bibitem{LOFF} M. G. Alford, J. A. Bowers and K. Rajagopal, Phys. Rev. \textbf{D63} (2001) 074016;
J. A. Bowers, J. Kundu, K. Rajagopal and E. Shuster, Phys. Rev. \textbf{D64} (2001) 014024;
J. A. Bowers and K. Rajagopal, Phys. Rev. \textbf{D66} (2002) 065002.
\bibitem{overhauser} B.-Y. Park, M. Rho, A. Wirzba and I. Zahed, Phys. Rev \textbf{D62} (2000) 034015;
P. Jaikumar and I. Zahed, Phys. Rev \textbf{D64} (2001) 014035.
\bibitem{CDW} F. Dautry, E. M. Nyman, Nucl. Phys. {\bf A319} (1979) 323;
T. Tatsumi, Prog. Theor. Phys. {\bf 63} (1980) 1252; M. Kutschera,
W. Broniowski and A. Kotlorz, Nucl. Phys. {\bf A516} (1990) 566; W. Broniowski,
M. Sadzikowski, Phys. Lett. {\bf B488} (2000) 63, M. Sadzikowski, Phys. Lett. \textbf{B553} (2003) 45;
E. Nakano and T. Tatsumi, Phys. Rev. \textbf{D71} (2005) 114006;
B. Bringoltz, JHEP 0703:016,2007; D. Nickel, Phys. Rev. \textbf{D80} (2009) 074025,
T. Partyka, M. Sadzikowski, J. Phys. \textbf{G36} (2009) 025004;
S. Maedan, Prog. Theor. Phys. \textbf{123} (2010)285.
\bibitem{MS} M. Sadzikowski, Phys. Lett. \textbf{B642} (2006) 238.
\bibitem{kojo} T. Kojo, Y. Hidaka, L. McLerran and R. D. Pisarski, Nucl. Phys. \textbf{A843} (2010) 37;
T. Kojo, R. D. Pisarski and A. M. Tsvelik, e-Print: arXive: hep-ph, 1007.0248 (2010).
\bibitem{klevansky} S. P. Klevansky, Rev. Mod. Phys. \textbf{64} (1992) 64.
\bibitem{dumitru} A. Dumitru, R. D. Pisarski and D. Zschiesche, Phys. Rev. \textbf{D72} (2005) 065008.
\bibitem{Hell} S. R\"o{\ss}ner, T. Hell, C. Ratti and W. Weise, Nucl. Phys. \textbf{A814} (2008) 118.
\bibitem{sasaki} C. Sasaki, B. Friman, K. Redlich, Int.J.Mod.Phys. \textbf{E16} (2007) 2319.
\bibitem{schaefer} B. J. Schaefer, J. M. Pawlowski and J. Wambach, Phys. Rev. \textbf{D76} (2007) 074023.
\bibitem{fukushima2} K. Fukushima, e-Print: arXive: hep-ph,  1006.2596 (2010)
\bibitem{lattice} F. Karsch and E. Laermann, Phys. Rev. \textbf{D50} (1994) 6954.
\bibitem{Blaschke} D. Blaschke, H. Grigorian, A. Khalatyan, D.N. Voskresensky, Nucl. Phys. \textbf{B141} (2005) 137.
\bibitem{Andronic} A. Andronic et al., Nucl. Phys. \textbf{A837} (2010) 65.
\bibitem{Nickel} D. Nickel, Phys. Rev. Lett. \textbf{103} (2009) 072301.
\bibitem{partyka} Some of the results can be already found in T. Partyka, e-Print: arXiv: hep-ph, 1005.5688 (2010).
\end{thebibliography}
\end{document}